\documentclass[preprint,12pt]{elsarticle}

\usepackage{epstopdf}
\usepackage{amsmath,amssymb}
\usepackage{amsmath, amsthm, amssymb}
\usepackage{graphicx}
\usepackage{dcolumn}
\usepackage{bm}
\usepackage{ifpdf}
\usepackage{multirow}
\usepackage{multirow}
\usepackage{array}
\usepackage{amsmath}
\usepackage{color,soul}
\usepackage{xcolor}

\journal{}

\begin{document}

\begin{frontmatter}


\title{Coupled mechanical resonators with broken Lorentz reciprocity for sensor applications}


\author{Lijie Li}

\address{College of Engineering, Swansea University, Bay Campus, Swansea, SA1 8EN, UK. Email: L.Li@swansea.ac.uk}

\begin{abstract}
Having simultaneously a high quality factor (i.e. a narrow resonant band) and a shorter decay time between the resonating system and the external sources (i.e. a wide resonant band) is a desirable characteristic for mechanical resonators, which however has been regarded as contradictory. This has been known as the limit of Lorentz reciprocity. We explore a configuration to achieve this desired characteristic within the mechanical regime. The configuration consists of a pair of mechanical resonators coupled together through their connecting part. One of them is encapsulated in a vacuum environment, and the other is left in the normal ambient condition. Numerical model of this configuration shows clearly the advantages such as: (a), sensitivity to the change of resonant frequency is greatly improved (the product of bandwidth $\Delta \omega$ and the decay time $\Delta t$ has increased  at least two orders of magnitude); (b), the value of $\Delta \omega \cdot \Delta t$ can be adjusted through the coupling stiffness. 
\end{abstract}


\end{frontmatter}


\section*{Introduction}

Mechanical resonators have been used as one of key parts of many devices such as in communications devices \cite{RN130}, sensors \cite{RN139, RN135, RN125, RN140, RN127, RN136, RN132, RN133} and energy harvesting devices \cite{RN134}. Like the manufacturing technologies have been advancing the IC (integrated circuit) industry, nano-mechanical devices have emerged with much improved performances \cite{RN139, RN135, RN125, RN126, RN138, RN137}. These nano-mechanical resonators are often used for investigating quantum mechanics aspects in conjunction with optics \cite{RN128, RN131}. Quality ($Q$) factor  characterizing the sharpness of the resonant band of mechanical resonators should be critically considered in the design process of those resonating devices. $Q$ is defined as $Q=\omega_0/\Delta \omega$, where $\omega_0$ is the fundamental frequency at which the first peak amplitude appears and $\Delta \omega$ is the frequency band, which is usually extracted from the half-power bandwidth of the amplitude-frequency curve. $Q$ is also expressed as $Q=2\pi \frac{E_{max}}{\Delta E}$ \cite{Qbook}, where $E_{max}$ is the maximum energy stored, and $\Delta E$ is the energy dissipated per cycle. Rewrite the expression of $Q$ leads to $Q\Delta E/E_{max} =2\pi$, which indicates that the product of $Q$ factor and energy loss rate is a constant. Directly from the definition of the quality factor\cite{RN123}, higher $Q$ implies a narrower band, and vice versa. A narrower resonant band indicates that the resonator has better capability to conserve the mechanical energy, i.e. a longer decay time $\Delta t$ (can also be understood as the decay time dissipating mechanical energy to the ambient), meanwhile a slower response to external excitation sources, i.e. taking longer time for the resonator to stabilize. Hence $\Delta t$ is reversely proportional to $\Delta E$, $\Delta t \propto 1/\Delta E$. As a contrast, a resonating device experiencing a wider resonant band will have a much faster response to external sources and a shorter $\Delta t$ (a shorter stabilization time and being able to respond to external sources with a much wider frequency band). This phenomenon has been well recognized and known as Lorentz reciprocity, written as $\Delta t \cdot \Delta \omega \sim C$ ($C$ is a constant, which was noted as 2$\pi$ \cite{RN122}). The shape of frequency response of a linear time invariant resonator resembles a Lorentzian, a characteristic of the exponential decay in the time domain, with the amplitude-time relation being mathematically written as $\cos(\omega_0 t)\cdot e^{-(C/2)\Gamma t}$, and the amplitude-frequency relation is written as $A(\omega)\propto \frac{(\Gamma/2)^2}{(\omega-\omega_0)^2+(\Gamma/2)^2}$, where $\Gamma$ is the energy loss rate/decay rate, inversely proportional to the $\Delta t$, $\Gamma \propto C/\Delta t$. For a linear, time-invariant resonating system, the Lorentzian explains $(\omega_0+\Gamma/2)-(\omega_0-\Gamma/2)=\Delta \omega=\Gamma$, hence $\Delta t \cdot \Delta \omega \sim C$. This constant sets the upper limit of the performance of all types of resonators including electronic and mechanical resonators, which means that there is a trade-off between the response time and the $Q$ factor (normally contributing to the sensitivity of the resonant frequency change). This rule states that a short response time and a high $Q$ cannot exist simultaneously. From the perspective of designing high performance micro/nano mechanical resonators, high frequency sensitivity (high $Q$) is required \cite{RN137, RN138}, as well as the device should be capable of responding to external sources with a wider frequency band, i.e. the device can be excited even if the excitation frequency deviates significantly from the device resonant frequency. 

Reference \cite{RN122} theoretically proposed a method to break the Lorentz reciprocity for a semiconductor heterostructure, where the product $\Delta t \cdot \Delta \omega$ was increased by several orders of magnitude. In this work, for the first time, we demonstrate the broken Lorentz reciprocity for a mechanical resonator. The traditional design of a coupled mechanical resonator depicted in Figure 1a shows that the resonator A is driven by an external source and the resonator B is driven by A. They both have a similar $Q$, and the rule of Lorentz reciprocity applies to each individual of them. It is seen that the driving energy $E_{drive}=E_{couple}+E_{damp}+E_{damp\textunderscore2}+E_A+E_B$, where $E_A$ and $E_B$ are vibration energies of A and B, $E_{drive}$ denotes to the external energy fed into A, $E_{couple}$ represents the mechanical energy exchange between A and B, and $E_{damp}$ and $E_{damp\textunderscore2}$ stand for the amount of energies dissipated to the surrounding environment through air damping. Here thermal energy loss (increase of the vibration amplitude of atoms in the crystal lattice) is neglected. Figure 1b sketches out a design that allows the resonating system to exhibit a broken Lorentz reciprocity.  

\begin{figure}[!t]
\centering
\begin{tabular}{c}
\includegraphics[width=4.6in]{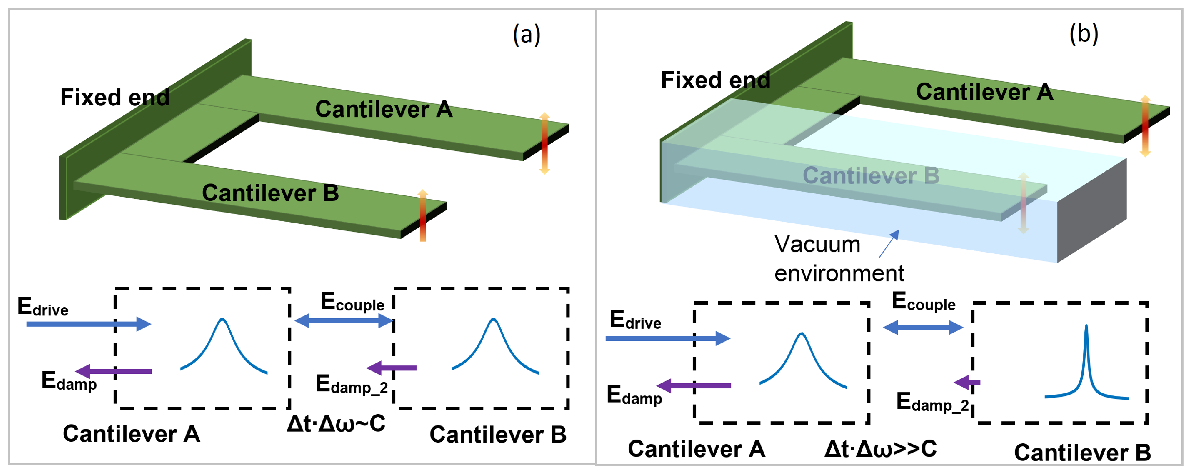} \\
\includegraphics[width=3.6in]{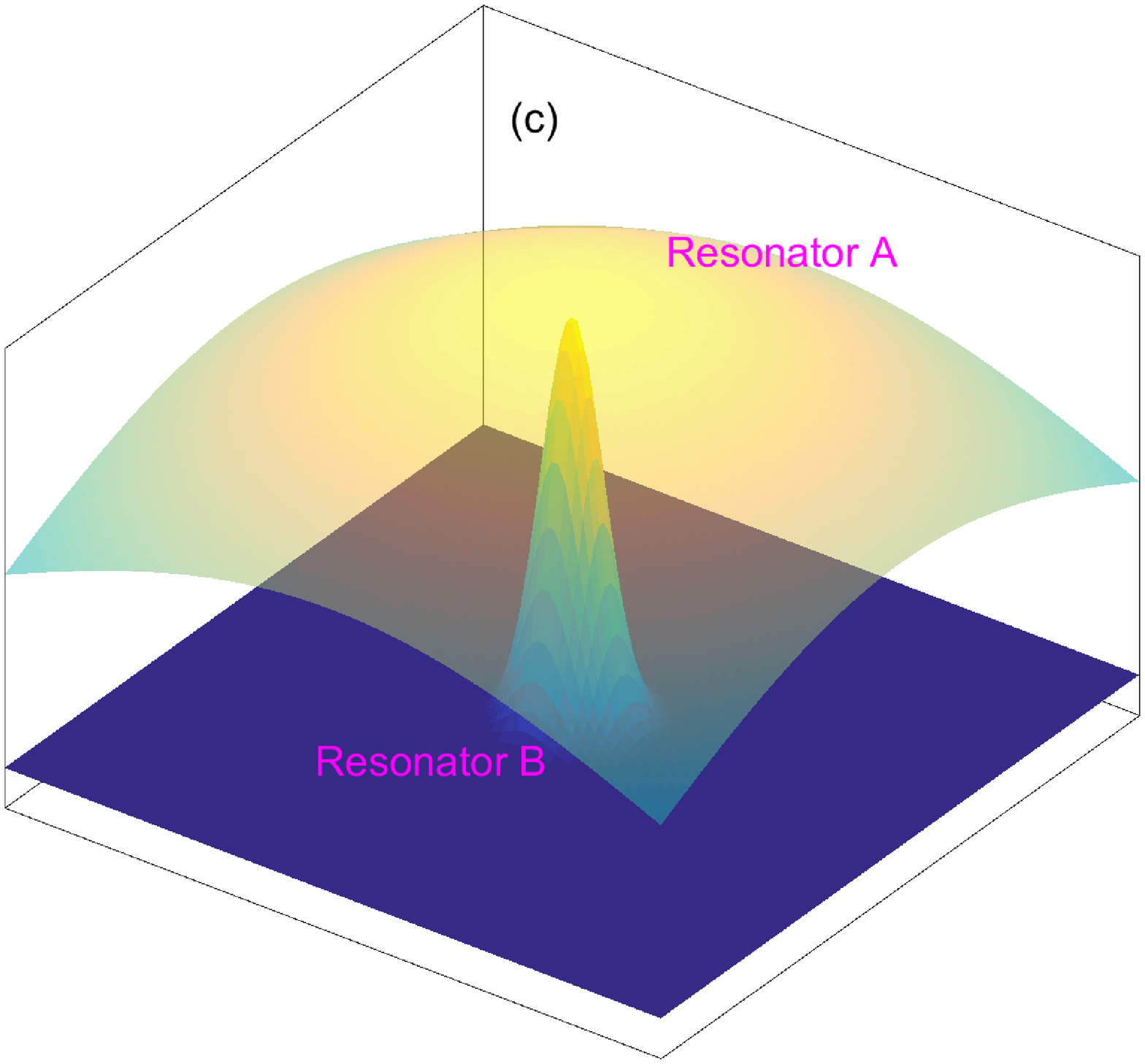}
\end{tabular}
\caption{Schematic of the general coupled mechanical resonators obeying Lorentz reciprocity rule, where two cantilevers have similar $Q$ factors (a). (b), cantilever A driven by an external source is in a general environment having a relatively large $\Delta \omega$, whereas cantilever B is in a vacuum environment exhibiting a smaller $\Delta\omega$. The system is able to respond mechanical excitations with a wide range of excitations as well as display a high frequency sensitivity with the cantilever B. $\Delta\omega_A\cdot\Delta t_B$ is several orders of magnitude higher than the constant value of traditional systems. (c), Schematic sketch of two coupled resonators having a wider Lorentz (A) and a sharper Lorentz (B). }
\label{fig_1}
\end{figure}

\section*{Coupled resonator with broken Lorentz reciprocity}

Bearing the aim of breaking the Lorentz reciprocity for the mechanical resonator, we structure two cantilevers coupled together (Figure 1b). A is in the general ambient and B is sealed in a vacuum environment, hence $c_A \gg c_B$, where $c_{A,B}$ are the damping coefficients caused by the drag force of air molecules acting on the resonators. Using this design, A is activated by an external excitation, which drives B through the connection part between them. Treating this coupled pair of cantilevers as a resonating system, energy from the driving source enters to A with the response time $\Delta t_A$, then a portion of vibration energy of A is coupled to B. Because the damping coefficient of B is much smaller, the energy dissipating to the surrounding environment through air damping is significantly slowed down, hence a much longer decay time $\Delta t_B \gg \Delta t_A$. 

The mechanical coupling factor $k_c$ acts on the displacement $x$, $(k_i+k_c)x_i$, $i$ denotes A or B, and the damping factor acts purely on the first derivative of $x$, $c_i \dot{x}_i$. Therefore the decouple of the air damping and the energy exchange between the two resonators leads to the broken Lorentz reciprocity for this resonating system. Qualitatively in a mechanical resonating system ($m\ddot{x}+ c\dot{x}+kx = a\cos(\omega t)$), a general steady state solution for an underdamped system is \\ $x=\frac{ac\omega}{(k-m\omega^2)^2+(c\omega)^2}\sin(\omega t)+\frac{a(k-m\omega^2)}{(k-m\omega^2)^2+(c\omega)^2}\cos(\omega t)$. It is clear that the time-symmetry ($T:t\mid \longrightarrow -t$) is damaged by the non-zero damping coefficient $c$.

To validate the concept, the lumped model for this linear, time-invariant (LTI) system can be expressed using the well established classic Mass-Spring-Damper system \cite{RN124, RN127}, that is
\begin{align}
\begin{split}
&m_A\ddot{x}_A+c_A \dot{x}_A+(k_A+k_c)x_A-k_cx_B =a\cos(\Omega t)\\
&m_B\ddot{x}_B+c_B \dot{x}_B-k_c x_A+(k_B+k_c)x_B =0
\end{split}
\end{align}
\noindent where $m_{A,B}$ represents the effective mass of the resonator, and $k_{A,B}$, $k_c$ are stiffness of individual cantilever and coupling stiffness respectively. There are several ways to arrive at solutions for this LTI system, Here the state-space model of the system is used, the state-space form of the system is written as

\begin{align}
\begin{split}
&\lbrace \dot{x} \rbrace = [\alpha] \lbrace x \rbrace + [\beta] \lbrace f_D \rbrace\\
&\lbrace y \rbrace = \lbrace x \rbrace
\end{split}
\end{align}

\noindent where $\lbrace x \rbrace$ is the state vector, $\lbrace y \rbrace$ denotes the output vector. $[\alpha]$ and $[\beta]$ are state and input matrix respectively, output matrix is a $4\times 4$ unit matrix and the feed-through force is 0. Re-write equation (1) in the form of the equation (2), it is
 
\begin{align}
\begin{split}
&\lbrace \dot{x} \rbrace =
\left[\begin{array}{c} \dot{x}_A \\ \ddot{x}_A \\ \dot{x}_B \\ \ddot{x}_B \end{array}\right], \lbrace x \rbrace =
\left[\begin{array}{c} x_A \\ \dot{x}_A \\ x_B \\ \dot{x}_B \end{array}\right]  
\\
&[\alpha]=
\left[\begin{array}{cccc} 0 & 1 & 0 & 0 \\ -\frac{k_A+k_c}{m_A} & -\frac{c_A}{m_A} & \frac{k_c}{m_A} & 0 \\ 0 & 0 & 0 & 1 \\ \frac{k_c}{m_B} & 0 & -\frac{k_B+k_c}{m_B} & -\frac{c_B}{m_B} \end{array}\right]
\\
&[\beta]=
\left[\begin{array}{c} 0 \\ \frac{1}{m_A} \\ 0 \\ 0 \end{array}\right]
\end{split}
\end{align}

 The above equations can be easily solved numerically. Parameters with reduced units are used in the lumped model, and they are: $m_A=m_B=1$, $k_A=k_B=1$, $c_A=0.5$, and $k_c=0.005$. At first, we calculate the frequency response of the resonating system when B is in the vacuum with various vacuum levels, while the damping coefficient of A remains unchanged. It is shown in Figure 2a that there is no change to the A in terms of amplitude and $Q$ factor. The $Q$ of B has changed in a huge amount when $c_B$ decreases from 0.5 (unless noted the units are taken as arbitrary units (a.u.) in following sections) to 0.005, and the amplitude of B has increased in the meantime. To clearly display the frequency response, i.e. $Q$ factor, we normalize the amplitudes of B, which is shown in Figure 2b.

\begin{figure}[!t]
\centering
\includegraphics[width=4.4in]{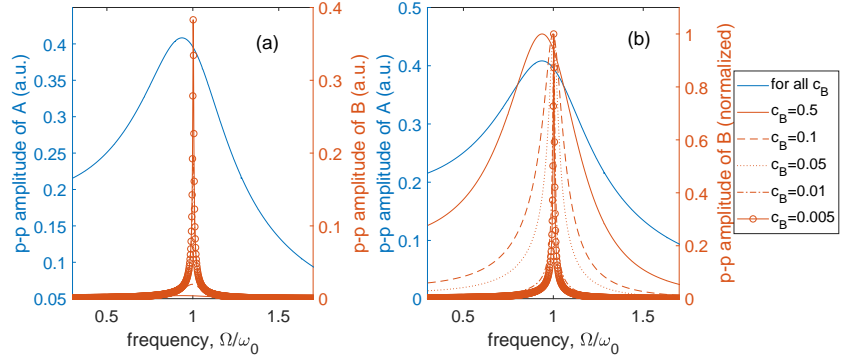}
\caption{(a), frequency response to the external source ($\sim\cos(\Omega t)$) of the coupled resonating system when B is in various vacuum levels. (b), resonating amplitude of B is normalized to show clearly the change of $Q$ in relation to the damping coefficient.}
\label{fig_2}
\end{figure}

Figure 3 shows the results of the product of the decay time and the half-power bandwidth of the system. It is shown when the vibration energy is transmitted in the direction from A to B, $\Delta \omega_A \times \Delta t_B$ has increased more than $10^2$ times, and $\Delta \omega_B \times \Delta t_A$ has reduced to the minimum of around 0.01 as the vacuum level surrounding B increases from 2 to 200. When the transmission direction is reversed (B is excited by the external source, and A is driven by B), both products ($\Delta \omega_B \times \Delta t_A$ and $\Delta \omega_A \times \Delta t_B$ )are calculated to be the same constant. This direction determined imbalance demonstrates the key characteristics of the nonreciprocal systems.\cite{Nonr}

\begin{figure}[!t]
\centering
\includegraphics[width=3.8in]{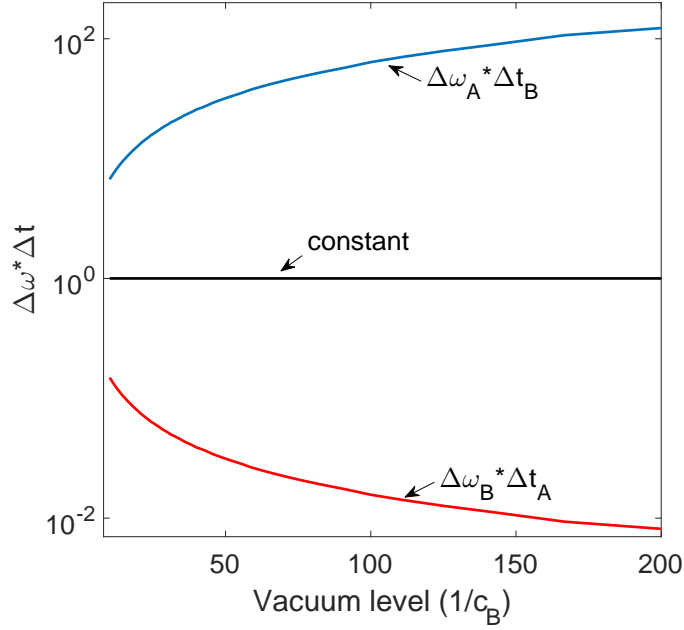}
\caption{Vacuum level ($1/c_B$) increases up to 200, $\Delta \omega_A \times \Delta t_B$ has been calculated to be 100 times more than the value set by the Lorentz reciprocity rule for the system where two resonators are in the same damping environment (central line). $\Delta \omega_B \times \Delta t_A$ is calculated to be much smaller than the constant. When the input and output are interchanged, i.e. B is master resonator that is excited by an external source, and A is the slave resonator that is driven by B, both $\Delta \omega_A \times \Delta t_B$ and $\Delta \omega_B \times \Delta t_A$ are calculated to be the unit constant. This transmission direction dependent behaviour resembles the key characteristics of the nonreciprocal devices.}
\label{fig_3}
\end{figure}

Further investigation on the impact of the coupling factor ($k_c$) between two resonators on the $\Delta t_B$ has been performed with various values of vacuum levels for B ($1/c_B$). It is found in Figure 4 that the decay time $\Delta t_B$ ($\propto Q$) increases as the $k_c$ reduces and the vacuum level ($\sim \frac{1}{c_B}$) increases. Understood from previous studies, lowering the damping coefficient increases the $Q$ factor. In the same time $\Delta t_B$ is also dependent on the mechanical coupling factor. Larger $k_c$ will enhance the energy exchange between two resonators, subsequently merging two $Q$s, from which the direct observation is the bandwidth of the resonator A reduces and that of B increases. Therefore in order to achieve a higher $\Delta t \Delta \omega$, a weaker coupling ($k_c<<k_{A,B}$) between two resonators should be chosen.  

Statistical Energy Analysis (SEA) \cite{SEA1, SEA2} can also be used to study this coupled system. Based on the SEA theory, the classical power flow relationship between coupled resonators is $P = \beta (E_A-E_B)$, where $E_A$ and $E_B$ are vibrational energies of two resonators. The factor $\beta$ given in equation (2.2) of \cite{SEA2} reduces to $\beta = k_c^2/(c_A+c_B)$ with $m_A=m_B=\omega_A=\omega_B=1$. As the half-power bandwidth is driven by the internal damping, $\Delta \omega_A=c_A/m_A=c_A$. The decay time is driven by the internal damping and loss by energy exchange, $\Delta t_B=2\pi/(\Delta \omega_B + \beta)$. Hence $\Delta \omega_A \times \Delta t_B = 2\pi c_A/(c_B+\frac{k_c^2}{c_A+c_B})$ and  $\Delta \omega_B \times \Delta t_A = 2\pi c_B/(c_A+\frac{k_c^2}{c_A+c_B})$. These two expressions allow to produce similar results in Figures 3 and 4, and to discuss the case of weak coupling and other asymptotic behaviours.

\begin{figure}[!t]
\centering
\includegraphics[width=3.8in]{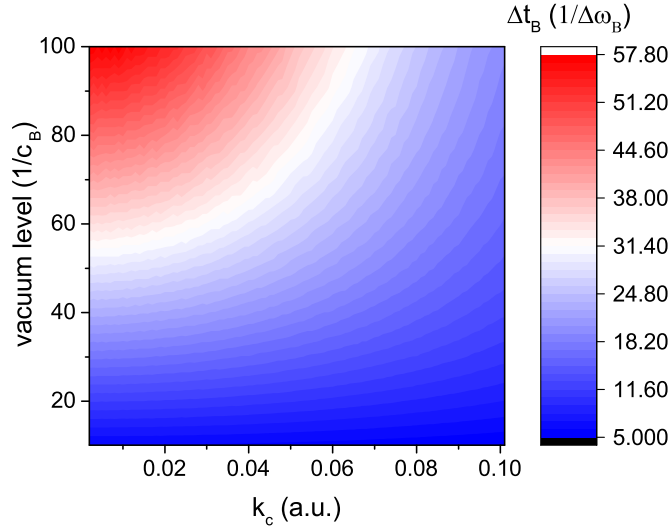}
\caption{$Q$ factor ($\sim$ decay time $\Delta t_B$) of B as functions of coupling factor $k_c$ and $1/c_B$.}
\label{fig_4}
\end{figure}

\begin{figure}[!t]
\centering
\begin{tabular}{c}
\includegraphics[width=3.6in]{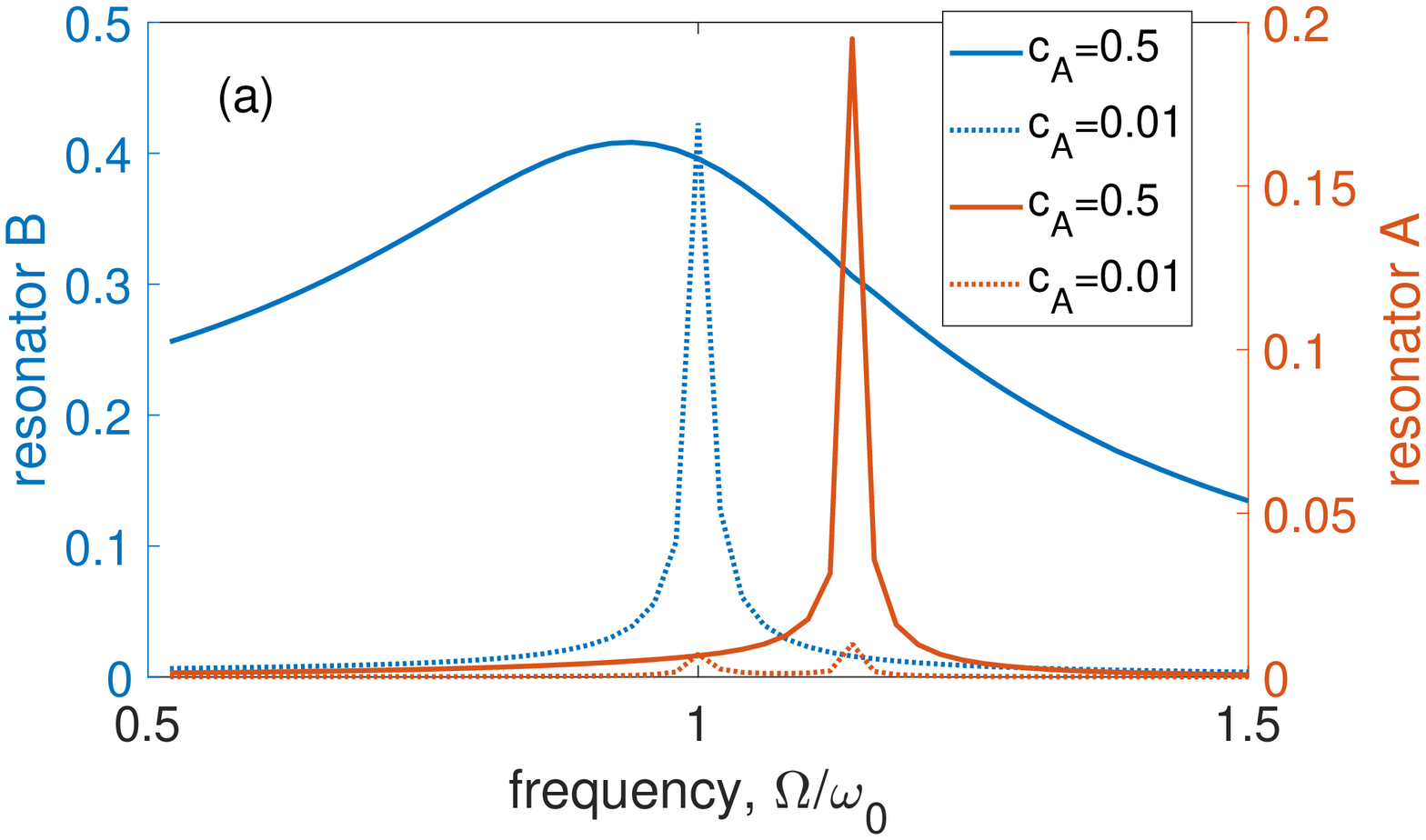}\\
\includegraphics[width=3.8in]{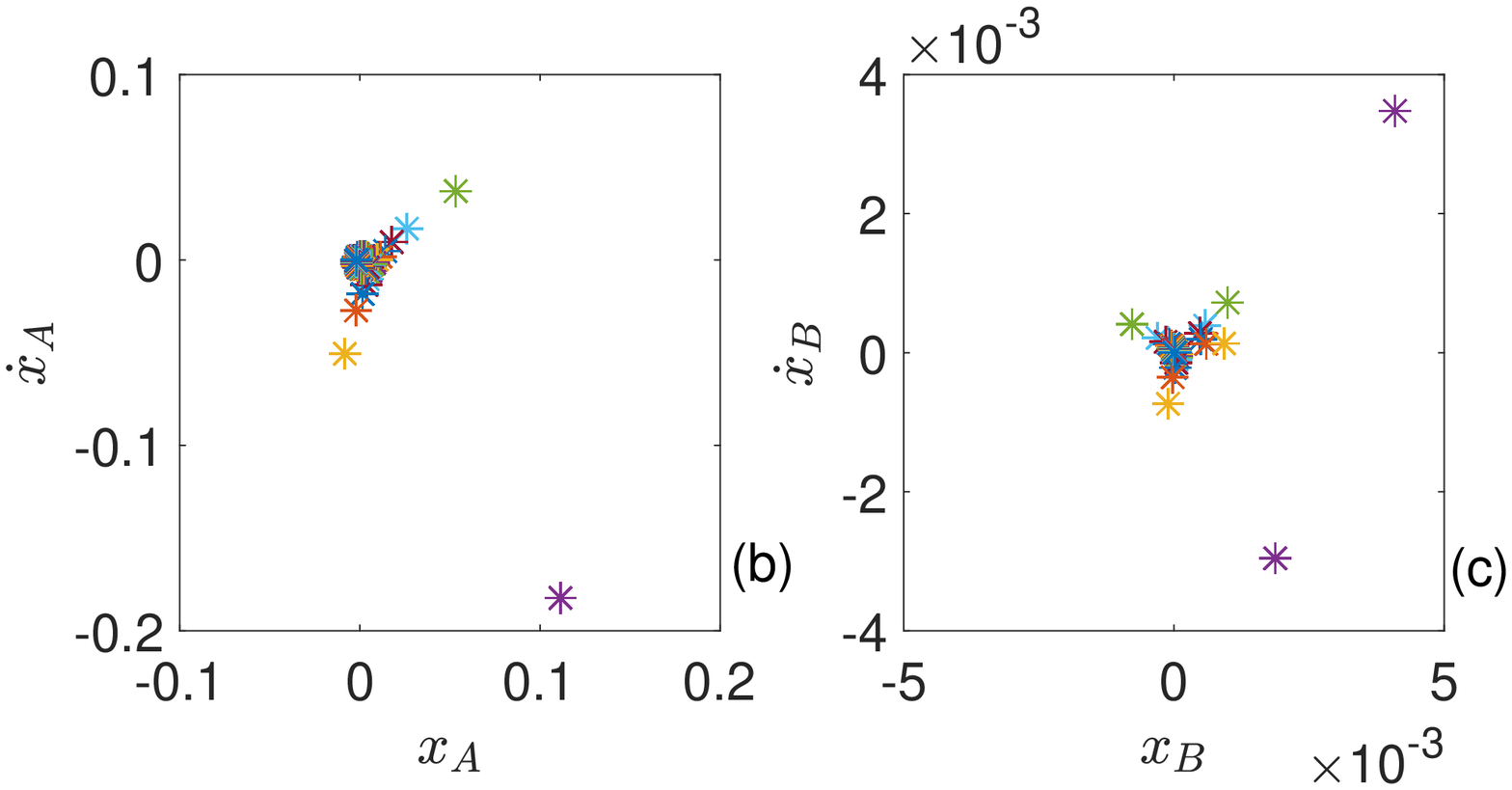}
\end{tabular}
\caption{(a), When both resonators are in the vacuum condition, e.g. $c_A=0.01$ and $c_B=0.005$ (dotted lines), the response of the resonator B is very small compared with resonator A is in general condition ($c_A=0.5$, solid lines), where the response of the resonator B is approximately 20 times higher. Here the peak amplitude of the resonator A remains similar. (b), Poincare Map of the resonator A at $c_A=0.01, c_B=0.005$. (c) Poincare Map of the resonator B at $c_A=0.01, c_B=0.005$.}
\label{Fig_5}
\end{figure}

\begin{figure}[!t]
\centering
\includegraphics[width=4.6in]{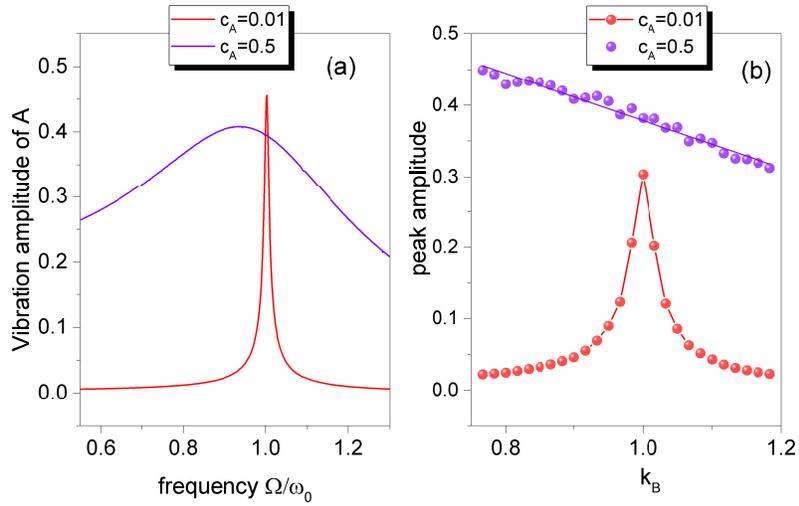}
\caption{(a), Frequency response for the resonator A at two damping conditions. (b), Response peak of the resonator B at two damping conditions in relation to its stiffness.}
\label{fig_6}
\end{figure}

\begin{figure}[!t]
\centering
\includegraphics[width=3.3in]{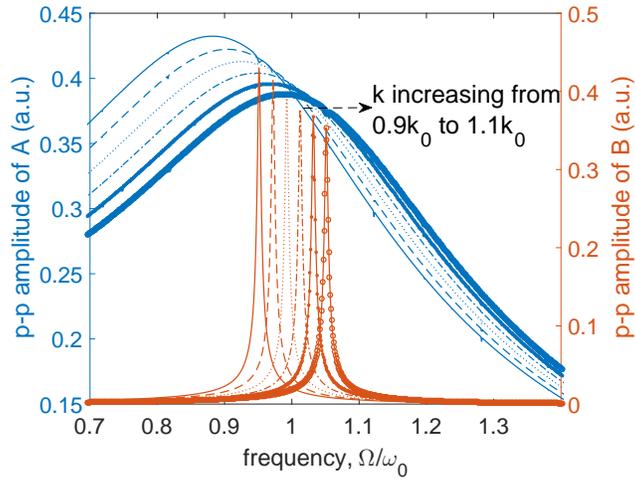}
\caption{Demonstration of the configuration exhibiting broken Lorentz reciprocity in a sensor application.}
\label{fig_7}
\end{figure}

\begin{figure}[!t]
\centering
\begin{tabular}{c}
\includegraphics[width=4.6in]{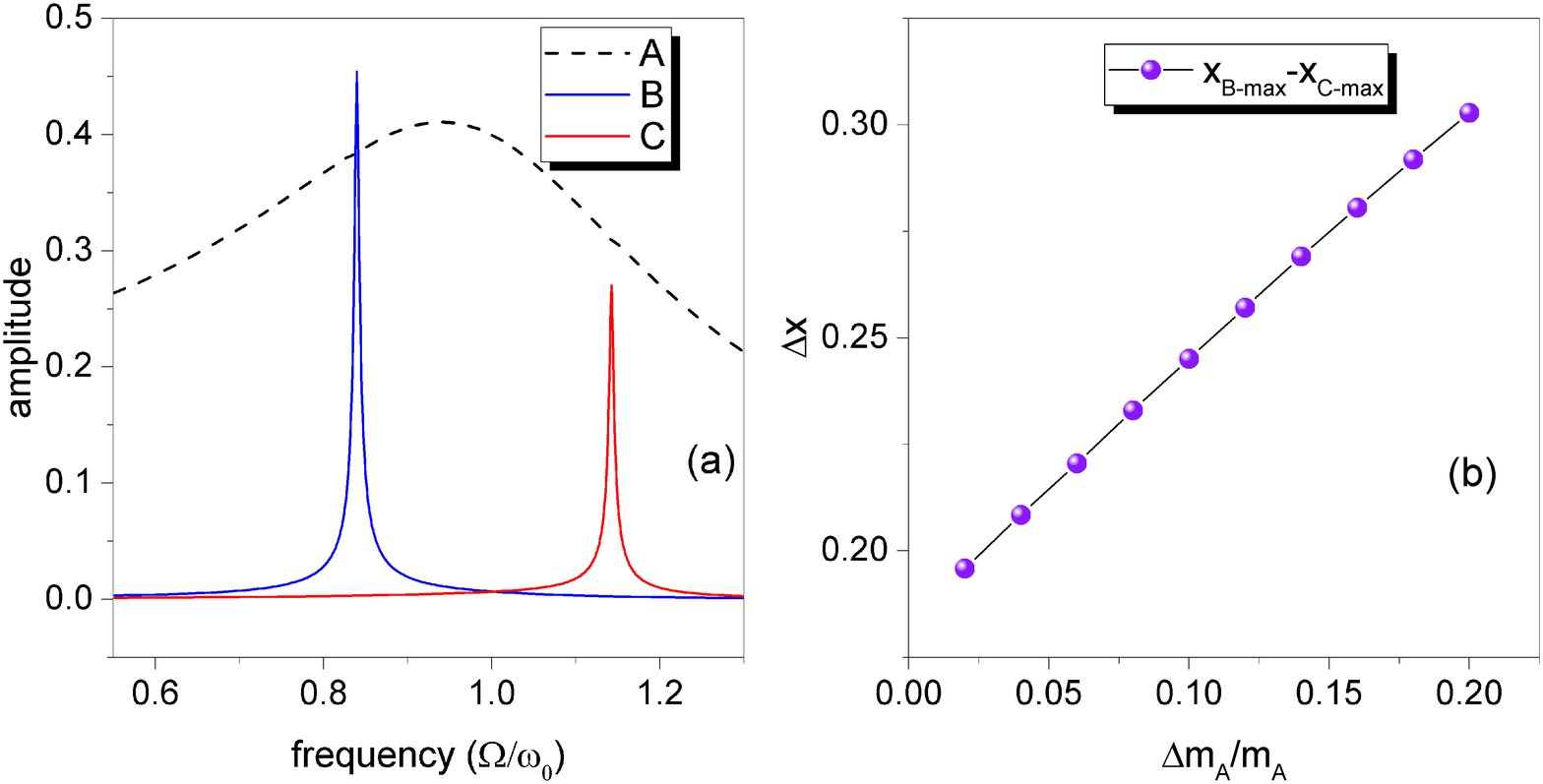} \\
\includegraphics[width=3.4in]{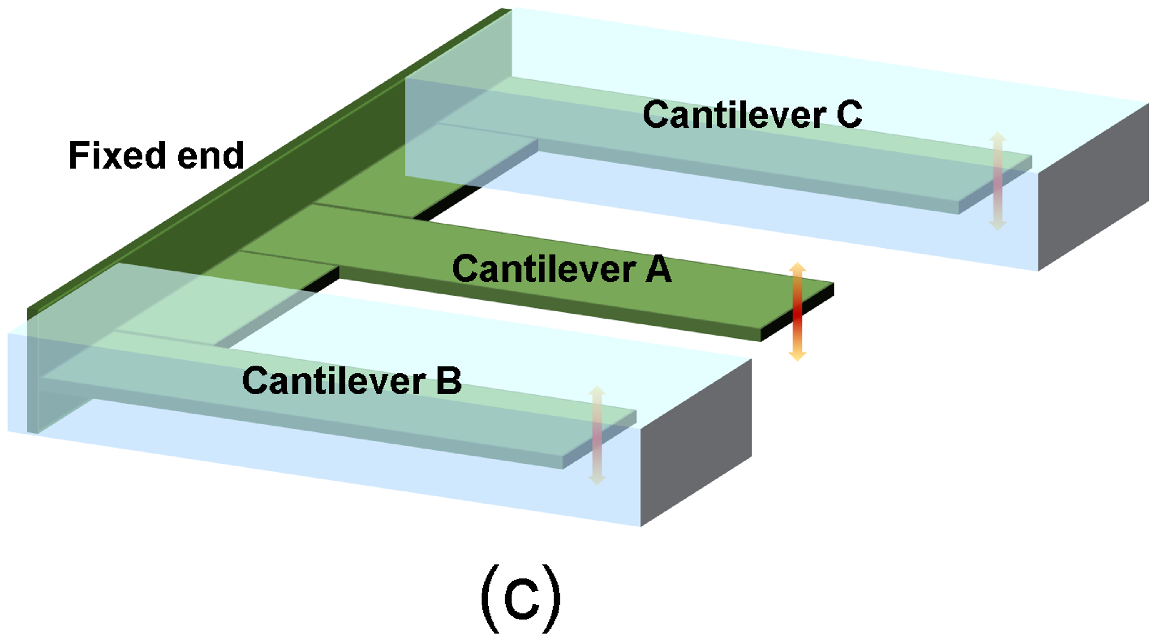}
\end{tabular}
\caption{(a), Frequency response of three coupled resonators. (b), $\Delta x$ vs. $\Delta m_A/m_A$. (c), Schematic of the coupled resonators with broken Lorentz reciprocity in a mass sensor design.}
\label{fig_8}
\end{figure}

In this symmetry-broken LTI system, the $\Delta t$ has been increased by enclosing the resonator B into a vacuum environment, meanwhile leaving the resonator A in a general environment to have relatively wide half-power bandwidth $\Delta \omega$. We also researched scenarios when the air damping of A changes as well. In Figure 5, results show the vibration amplitudes of A and B relating to the driving frequency at two conditions, $c_A=0.5$ and $c_A=0.01$, i.e. one stands for the general environment, and the other stands for the vacuum condition. When both A and B are in vacuum condition, it shows the disadvantage of the narrow bandwidth of A, which lowers the response of B (red dashed line). In the largely imbalanced damping conditions as proposed, very wide bandwidth of the resonator A results in much higher responses of resonator B (red solid line). In Figure 5, the stiffness of B has been shifted to $1.3k_0$ to offset from the peak of the resonator A. The simulation clearly demonstrates that the proposed LTI system with broken Lorentz symmetry benefits from the combination of both the wide bandwidth and high $Q$. It can be seen that there are two peaks for the response of B when A is in a vacuum condition, which should not be confused as it having multiple periods. Poincare maps of the resonators A and B (Figures 5b and 5c) elucidate that there is no multiple-period generation on both the A and B, although phases of them do vary for different driving frequencies. Further validation has been provided through simulation results in Figure 6, where it shows that at two conditions ($c_A=0.01$ $c_A=0.5$), the frequency response (vibration amplitude vs. driving frequency) of the resonator A (Figure 6a), and the peak response of resonator B (Figure 6b) in which the stiffness of the B ($k_B$) has linearly varied from 0.7$k_0$ to 1.2$k_0$. Results directly show that the peak amplitude of B is more uniform for the scenario of A being in the general environment ($c_A=0.5$) than that of A being in the vacuum, which implies that the Lorentz reciprocity limit has been overcome with this largely unbalanced damping configuration.    

\section*{Sensor application}

To demonstrate the advantage of having larger $\Delta t \Delta \omega$, we apply the system to a sensor device. The proposed sensing system is able to detect change of material properties, such as stiffness, which could be the consequence of the change of certain environmental parameters e.g. temperature \cite{RN143, RN144}. It is supposed that the stiffness of both A and B has changed from 0.9$k_0$ to 1.1$k_0$ ($k_0$ being the initial stiffness), simulation results show (Figure 7) that it is hard to detect the change precisely using the resonator A, while the resolution of using resonator B is much higher.     

Another application of this concept can be a mass sensor that consists of three coupled resonators of which two have lower damping coefficients, as depicted in Figure 8c. Similar simulation process has been followed to model the device. Here we set the stiffness of B and C to $0.7k_0$ and $1.3k_0$ respectively to distinguish two resonant peaks. In Figure 8a, the frequency responses of three resonators are displayed, where curve A is for the resonator A having relatively large air damping, the other two are in a vacuum environment (higher $Q$). Any added mass on the resonator A will result in direct resonant frequency change of A itself, however because its $Q$ is small, the resolution of purely replying on extracting the frequency change of A is low. Taking use of the strategy to have much higher $\Delta t \Delta \omega$, we can detect the difference between two resonators in the vacuum, which will benefit from a much higher resolution. The $\Delta x$ (maximum amplitude of the resonator B - maximum amplitude of the resonator C) is linearly increasing from $\sim$ 0.2 to $\sim$ 0.3 with the added mass $\Delta m_A/m_A$ from 2$\%$ to 20$\%$ of $m_0$. (Figure 8b). It corresponds to $\Delta x$ of $\sim$ 1.2$\%$ per 1$\%$ mass change. These coupled resonators (two, three or more) have to be considered as a whole system. The excitation (input) is on the A whose response time is short due to its large damping coefficient. The vibration energy is then coupled to B, and is dissipated to the environment in a much slower speed due to the smaller damping coefficient. It is true that a portion of vibration energy is coupled back to A, in which case it is a classical reciprocity system. The nonreciprocity exists and only valid for the scenario where the portion of vibration energy dissipated from the slave resonators residing in the lower damping environment. It is the modulation of the damping factor $\Delta c$ acting on the first order derivative of the $x$ to the different part of the resonating system that leads to the broken symmetry.

\section*{Conclusion}

To summarize, a concept that can essentially break the limitation set by the Lorentz reciprocity in mechanical resonating systems has been proposed. An example of using two coupled resonators has been analysed, and results explicitly demonstrated advantages. In addition, two examples of applying the concept in sensors have been modelled. Apart from the applications described above, inertia devices such as accelerometers, gyroscope, as well as energy harvesting devices will also be beneficiaries. It is noted that although the concept has been demonstrated using the form of cantilever, other mechanical resonating structures also work with this strategy.


\section*{Additional information}

\textbf{Competing interests}\\ 
The author declares no competing interests.

\end{document}